\documentclass[aps,pre,showpacs, twocolumn]{revtex4}

\usepackage{graphicx}
\usepackage{amssymb}
\usepackage{amsmath}
\usepackage{colordvi}
\usepackage{color}

\newcommand{\PT}{{\cal PT}}

\newcommand{\cF}{{\cal F}}
\newcommand{\cK}{{\cal K}}
\newcommand{\cL}{{\cal L}}

\newcommand{\bp}{{\bf p}}

\newcommand{\tw}{\tilde{w}}

\newcommand{\tb}{\tilde{b}}

\newcommand {\Image}{\textrm{Im\ }}

\newcommand{\vep}{\varepsilon}

\begin{document}

\title{Small-amplitude nonlinear  modes  under the combined effect of the parabolic potential, nonlocality and $\PT$ symmetry}

\author{Dmitry A. Zezyulin, Vladimir V. Konotop}

\affiliation{Centro de F\'isica Te\'orica e Computacional,\\
Faculdade de Ci\^encias da Universidade de Lisboa, Campo Grande, Edif\'icio C8, Lisboa  P-1749-016, Portugal
}

\date{\today}

\begin{abstract}

We consider nonlinear modes of the nonlinear Schr\"odinger equation with a nonlocal nonlinearity and an additional $\PT$-symmetric parabolic potential. We show that there exists a set of continuous   families of nonlinear modes and study their linear stability in the limit of small nonlinearity. It is demonstrated that either $\PT$ symmetry or the nonlocality can be used to manage the stability of the small-amplitude nonlinear modes. The stability properties are also found to depend on the particular shape of the nonlocal kernel. Additional numerical simulations show that the   stability results remain valid not only for the  infinitesimally small nonlinear modes,  but also for the modes of finite amplitude.

\end{abstract}


\maketitle

\section{Introduction}
\label{intro}

Nonlocal nonlinearities are known to be of relevance for
numerous physical applications. In optical applications, nonlocal
nonlinearity emerges in description of beam propagation in thermal
media~\cite{Litvak66} or in atomic vapors~\cite{vapors}. Nonlocal nonlinearities have been shown to support optical solitons \cite{Tur85, KrolBangSolitons, XuKart05,Skupin06}  and to alter their properties, such as, for instance, stability \cite{Tur85}, the collapse dynamics  \cite{BangKrolCollapse}, or the interactions between solitons \cite{DSNonlocal}.

Nonlocality is also relevant in the theory of Bose-Einstein  condensates (BECs) \cite{PS} in view of the intrinsically nonlocal nature of the inter-atomic interactions
\cite{Goral,Sinha}.  Nonlocal nonlinearities are also inherent for plasma problems~\cite{Litvak78} and for
the theory on nematic crystals~\cite{Assanto03}.


In many relevant situations, nonlocal nonlinearity is accompanied by a certain potential which models a confining trap for a BEC   or a  refractive index modulation in an optical waveguide.  Presence of the external potential or  nonlocality enriches the variety of possible nonlinear modes allowing to observe the so-called higher (alias multi-pole) nonlinear modes \cite{XuKart05, nonlin_homo} apart from the best studied fundamental (alias ground state) nonlinear modes \cite{firstBEC, KrolBangSolitons}.   Nonlocal nonlinearities have been considered in combination with  the double-well   \cite{Kevr2011} and the periodic potentials \cite{Kart_per,Kart_per1, Cuevas}.
It looks however that a systematic study of nonlocal nonlinear modes in a parabolic potential has not been performed yet.
In our present study, we make a step towards filling  this  gap and  investigate  nonlocal nonlinear modes in a complex parabolic $\PT$-symmetric potential of the form $V(x) = x^2 - 2i\alpha x$, where $\alpha$ is a real coefficient. For $\alpha=0$ the potential is reduced to the real parabolic potential whose modes are fairly well studied in the local case \cite{nonlin_homo,AlZez}, but have attracted much less attention in the nonlocal context.

For a nonzero $\alpha$, the potential $V(x)$ is complex and  $\PT$ symmetric \cite{BendBoet}, which is reflected by the fact that the real and imaginary parts of  $V(x)$ are even and odd functions of $x$, respectively. $\PT$ symmetry 
can be implemented in diverse physical setting, including  optics \cite{Ruter}, atomic gases \cite{HHK}, and   BECs \cite{KGM08}.

Nonlinear modes  and solitons  have been studied in various  $\PT$-symmetric potentials with local \cite{Musslimani_1, PTfamiliesLocal, ZezKon12} and nonlocal  \cite{PTNonlocal} nonlinearities, but
the present work has a slightly peculiar focus. We are particularly interested in the examining the combined effect of the parabolic trapping, nonlocality and $\PT$ symmetry on the \textit{excited} (higher-order) nonlinear modes. The  focus of our study is additionally shifted towards the existence and stability  of  \textit{small-amplitude}  nonlinear modes, whose properties can be studied by means of asymptotic expansions describing bifurcations of nonlinear modes from the eigenstates of the underlying linear model.  The linear spectrum of the $\PT$-symmetric parabolic potential is available in the analytical, which makes it possible to reduce the stability problem   to   computing the real zeros of certain analytical expressions.  Our analysis shows that either the degree of the nonlocality or  the strength of the $\PT$ symmetry   can  be used to manage the stability of the excited modes. On the other hand, we reveal a dependence  of the stability properties on  the particular shape of the nonlocal kernel. We also touch upon the properties of nonlinear modes of finite amplitude and illustrate numerically that the   stability results hold  not only for the infinitesimally  small nonlinear modes but also remain valid  in a finite range of the  amplitudes of nonlinear modes.

The paper is organized as follows. In the next section, we present a model to be considered throughout the paper. In Sec.~\ref{sec:nonlin} we describe the linear properties of the model and introduce the families of small-amplitude nonlinear modes. In Sec.~\ref{sec:stab} we describe the methodology of the linear stability analysis. The main results of stability analysis are presented  in Sec.~\ref{sec:cons} and Sec.~\ref{sec:PT}   for the conservative and $\PT$-symmetric cases, respectively. In Sec.~\ref{sec:fam} we demonstrate numerically that the presented asymptotical stability results persist in a finite range of the amplitude of the modes. Concluding Sec.~\ref{sec:Concl} outlines the main results of the work.

\section{The models}
\label{sec:model}
We use the dimensionless nonlocal nonlinear Schr\"odinger equation with an additional parabolic potential and  nonlocal nonlinearity:
\begin{eqnarray}
\label{GPE}
i\frac{\partial q (x,z)}{\partial z} =-  \frac{\partial^2 q(x,z)}{\partial x^2}+(x^2-2i\alpha x)q(x,z)
\nonumber  \\
-\sigma q(x,z)\int_{-\infty}^{\infty}K(x-x')|q(x',z)|^2dx'.
\end{eqnarray}
The case $\alpha=0$ corresponds to the real parabolic potential, while nonzero $\alpha$ governs the strength of the $\PT$ symmetry. The system is subject to gain in the domain $x<0$ and is subject to losses for $x>0$.  In the optical context, equation (\ref{GPE}) models  beam guidance in a medium whose refractive index $n(x) =  n_r(x) + in_i(x)$ has parabolic modulation of the real part $n_r(x) = x^2$, and linear modulation of the imaginary part $n_i = 2\alpha x$. Equation (\ref{GPE})  with $\alpha=0$ also arises as a Gross-Pitaevskii equation \cite{PS} in the mean-field theory  the harmonically trapped a quasi-one-dimensional  BEC \cite{Goral,quasi-collapse, Cuevas}. In this case, $q(x,z)$ is a quasi-one-dimensional wavefunction \cite{Sinha}, and $z$ plays a role of  time.
%
%
Sign $\sigma=\pm 1$ corresponds to attractive (focusing)  and repulsive (defocusing) nonlinearities, respectively. Nonlocal properties of the model are described by the kernel $K(x)$  which is a positive function with normalization $\int K(x)dx=1$ (hereafter all the integrals are taken over the whole real axis).  Following to the previous studies on the subject \cite{XuKart05,Skupin06,Kart_per1,Kevr2011}, we  pay the primary attention to the  following  two kernels:

(i) The smooth Gaussian kernel
\begin{eqnarray}
\label{eq:gauss}
K(x)= \frac{1}{\sqrt{\pi}\lambda} e^{-x^2/\lambda^2}
\end{eqnarray}
where $\lambda>0$ governs strength of nonlocality.

(ii) Exponential kernel with the singularity at $x=0$: 
\begin{eqnarray}
\label{eq:exponential}
K(x)=\frac{1}{2\lambda} e^{-|x|/\lambda}.
\end{eqnarray}


(iii)  It will be also interesting to compare the obtained results with a slowly (algebraically) decaying kernel. While some studies (for example,  \cite{Kevr2011, Cuevas}) deal with the so-called cut-off (CO) kernel which decays as $|x|^{-3}$ as $x\to\pm \infty$, in our study we choose the algebraically decaying Lorentzian kernel (the Lorentzian)
\begin{eqnarray}
\label{eq:alg}
K(x) =  \left\{
\begin{array}{c}
\frac{\lambda}{\pi(\lambda^2 + x^2)} \mbox{\quad  if } \lambda > 0, \\[2mm]%
\delta(x) \mbox{\quad\quad if } \lambda = 0.
\end{array}
\right.
\end{eqnarray}
This choice is  motivated by the fact that the Lorentzian kernel (\ref{eq:alg}) allows for the relatively simple analytical evaluation of the convolution integrals   that arise in what follows. 

\section{Bifurcations of nonlinear modes}
\label{sec:nonlin}


The stationary nonlinear modes for Eq.~(\ref{GPE}) admit  the   representation $q(x,z) = e^{i\beta z}w(x)$, where $\beta$ is a real parameter. It is convenient to introduce the shifted parameter $b$ defined through the relation  $\beta = b - \alpha^2$, which allows to rewrite the equation for the  nonlinear modes in the following form:
\begin{eqnarray}
\label{stat}
w_{xx} - bw - (x-i\alpha)^2w \hspace{3cm}
\nonumber  \\
+\sigma w(x)\int_{-\infty}^{\infty}K(x-x')|w(x')|^2dx'=0,
\end{eqnarray}
subject to the zero boundary conditions at the infinity: $\lim_{x\to \pm \infty} w(x)=0$.

Let us first recall the properties of the underlying linear problem which formally corresponds to $\sigma=0$ in Eq.~(\ref{stat}).  The resulting equation which is considered as a linear eigenvalue problem for $b$ and $w(x)$, has a purely real and  discrete spectrum \cite{Kato,Znoj1,ZezKon12}. The eigenvalues $\tb_n$  can be listed   as  $\tb_n = -(2n+1)$, $n=0,1,\ldots$. The corresponding eigenfunctions   $\tw_n(x)$ read
\begin{eqnarray}
\label{eq:twn}
\tw_n(x)=c_nH_n(x-i\alpha)e^{-(x-i\alpha)^2/2},
\\
 c_n = \frac{e^{-\alpha^2/2}}{\sqrt{\sqrt{\pi}2^nn!G_n(-2\alpha^2)}}
\end{eqnarray}
where $H_n(x)$ and $G_n(x)$ are Hermite and Laguerre polynomials \cite{AS}, respectively. The choice of the  coefficients $c_n$ implies normalization  $\int |\tw_n|^2dx=1$. Each Hermite polynomial  $H_n(x)$ contains only powers of $x$ of the same parity as the number $n$. This implies that  in the conservative limit ($\alpha=0$) the eigenfunctions are real-valued   functions which have exactly $n$ zeros and have the same parity as the number $n$. For $\alpha\ne 0$, each eigenfunction $\tw_n(x)$ possesses  real even part and odd imaginary part for even $n$ and odd real part and even imaginary part for odd $n$.


Returning now to the nonlinear problem,  we look for bifurcations of small-amplitude nonlinear modes from the linear eigenstates. To this end, we introduce  the formal asymptotic expansions which  describe nonlinear modes $w_n(x)$  branching off
from the zero solution $w(x)\equiv 0$ at  the eigenvalues $\tb_n$ \cite{ZezKon12}:
 \begin{equation}
 \label{eq:expans}
w_n(x)=\vep\tw_n + \vep^3 w_n^{(3)} + \mathcal{O}(\vep^5), \quad b_n = \tb_n +
\sigma\vep^2b_n^{(2)} + o(\vep^2),
\end{equation}
where  $\vep\ll 1$ is a formal small real parameter (for the  rigorous analysis of the bifurcations of small-amplitude  modes see \cite{DS15}).

Since $\tw_n(x)$ are  normalized, the squared $L^2$-norm of the nonlinear modes $U=\int |w_n(x)|^2 dx$ (which is also associated with the total energy flow in the optical applications of the model) is approximately equal to $\vep^2$: $U \approx \vep^2$ for $\vep \ll 1$.  Substituting expansions (\ref{eq:expans}) into Eq.~(\ref{stat}) and collecting the terms of the $\vep^3$-order, one arrives at an inhomogeneous  linear differential equation with respect to the function $w_n^{(3)}$. The solvability condition for the latter equation yields  the expression for the coefficient  $b_n^{(2)}$:
\begin{equation}
\label{b2}
b_n^{(2)} = \frac{\int  \tw_n^2(x) \left (\int  K(x-x')|\tw_n(x')|^2 dx' \right) dx}{\int  \tw_n^2(x)dx}.
\end{equation}
For the conservative case ($\alpha=0$) the eigenfunctions $\tw_n(x)$ are real,  and hence the coefficients  $b_n^{(2)}$ are obviously   real too for any $n$.   For $\alpha \ne 0$,  the eigenfunctions $\tw_n(x)$ possess nontrivial real and imaginary parts  as described above. However,     parities of real and imaginary parts of $\tw_n(x)$  imply that for a  real and even kernel function $K(x)=K(-x)=K^*(x)$  the coefficients    $b_n^{(2)}$ are also real, which shows that continuous families of nonlinear modes can exist in the $\PT$-symmetric model with $\alpha>0$. The family with $n=0$ represents the fundamental nonlinear modes, and the families $n=1,2,\ldots$  contain  the higher-order modes.

\section{Linear stability of nonlinear modes}
\label{sec:stab}
\subsection{Statement of the problem}

Following to the the standard procedure of the linear stability analysis, we consider a perturbed nonlinear  mode $w_n(x)$ in the form
$q(x,z) = e^{i\beta z}[w_n(x) + u(x)e^{i\omega z} +
v^*(x)e^{-i\omega^*
    z}]$ which after linearization with respect to small functions
    $u(x)$ and $v(x)$ gives the
%
%
linear stability eigenvalue problem
\begin{equation}
    \label{BdG}
    {L}\, \bp  = \omega\, \bp , \quad
    {\bf p} =\left(
    \!
    \begin{array}{c}%
    u
    \\%
    v
    \end{array}
    \!\right),
\end{equation}
with the linear   operator $L$   given as
\begin{eqnarray*}
        \label{eq:L}
    {L} =
    \left(
    \!\!
    \begin{array}{cc}%
    A & \sigma w_n \cK_{w_n}
    \\[2mm]
    -\sigma w_n^*\cK_{w_n^{*}} &-A^\dag%
    \end{array}
    \!\!\right),
\end{eqnarray*}
where
\begin{eqnarray}
A &=& \cL + \sigma (\cK|w_n|^2)  + \sigma w_n \cK_{w^*_n},\\[0.5mm]%
A^\dag &=&  \cL^\dag + \sigma (\cK |w_n|^2)  + \sigma w_n^*\cK_{w_n},\\[0.5mm]
\cL &=&d^2/dx^2 - b - (x-i\alpha)^2,\\[0.5mm]%
\cL^\dag&=&d^2/dx^2 - b - (x+i\alpha)^2,
\end{eqnarray}
$\cK$ is the  convolution operator, i.e.,  $ \cK p =
\int  K(x-x') p(x')dx'$,  and   $\cK_p$  is the   operator of
``weighted'' convolution, i.e.,  $\cK_p\, q = \cK(p\,q) = \int
K(x-x')p(x')q(x')dx'$. 

The nonlinear mode $w_n(x)$ is said to be linearly  unstable if  operator $L$ has at least one  eigenvalue $\omega$ such that  $\Image \omega <0$. Otherwise,  $w_n(x)$ is linearly   stable.

\subsection{Linear stability of small-amplitude modes}

Let us now analyze the spectrum of the operator ${L}$  for small-amplitude  nonlinear modes  belonging to the $n$th family and situated  in the vicinity of the   bifurcation from the linear limit. For nonlinear modes of zero amplitude  (i.e.,
for $\vep=0$) the operator ${L}$ acquires the diagonal form
\begin{equation}
    {L} = \tilde{{L}}_n = \left(\begin{array}{cc}%
    {\cal L}_n & 0\\%
    0 & -{\cal L}_n^\dag%
    \end{array}\right),
\end{equation}
where 
\begin{eqnarray}
\label{eq:cL}  {\cal L}_n = \frac{d^2\ }{d x^2}  - \tb_n -
(x-i\alpha)^2,\\
{\cal L}_n^\dag = \frac{d^2\ }{d x^2}  - \tb_n -
(x+i\alpha)^2.
\end{eqnarray}
Since the spectrum of the operator $ {\cal L}_n + \tb_n$ is known from the discussion in Sec.~\ref{sec:nonlin}, it is easy to deduce that  the spectrum of the operator $\tilde{{L}}_n$
consists of two subsets. Eigenvalues and eigenvectors of the
first subset read $\omega^{(I)}_{n,k} = 2(n-k)$,
$\bp^{(I)}_{n,k} = (\tw_k(x), 0)^T$, $k=0,1,\ldots$. The second
sequence reads $\omega^{(II)}_{n,k} = -2(n-k)$, $\bp^{(II)}_{n,k}
= (0, \tw_k^*(x))^T$, $k=0,1,\ldots$.

Thus, operator $\tilde{{L}}_n$
has a double zero eigenvalue $\omega^{(I)}_{n,n} =
\omega^{(II)}_{n,n}=0$ which remains zero and double  when passing from
the linear limit $\vep=0$   to $\vep>0$ \cite{ZezKon12}. The operator $\tilde{{L}}_n$ also  has $2n$ double nonzero
eigenvalues: $\Omega_{n,k} = \omega^{(I)}_{n,k} =
\omega^{(II)}_{n,2n-k}$, where $k$ runs from 0 to $2n$ except for
$k=n$. Each double eigenvalue $\Omega_{n,k}$ is
semi-simple; the two corresponding linearly independent  eigenvectors  are given as
\begin{equation}
  \bp^{(I)}_{n,k}
= \left(\begin{array}{c}
\tw_k\\ 0
\end{array}\right) \mbox{\quad and\quad} \bp^{(II)}_{n,2n-k} = \left(
\begin{array}{c}
0\\ \tw_{2n-k}^*
\end{array}
\right).
\end{equation}

 Passing   from the linear limit $\vep=0$ to small nonzero  $\vep$, each
double eigenvalue $\Omega_{n,k}$ generically splits into two
simple eigenvalues which will  be either both real or complex
conjugated.  
 If each double
eigenvalue  splits into two real ones, then the small-amplitude
modes belonging to the $n$th family are  stable for both attractive and repulsive nonlinearities. \emph{Vice versa},
if at least one double eigenvalue gives birth to a pair of two
complex-conjugated simple eigenvalues, then the small-amplitude nonlinear modes of
the $n$th family are unstable.   Notice also that for $n=0$ no  double eigenvalues $\Omega_{n,k}$ exists.
Therefore the fundamental  solutions from    family $n=0$ are always stable in the small-amplitude limit.

It is easy to check that  the eigenvalue  $\Omega_{n,k}$  splits in exactly the same way
as the opposite double eigenvalue
$\Omega_{n,2n-k} = -\Omega_{n,k}$, i.e., if    $\Omega_{n,k}$  splits into two real eigenvalues then    $\Omega_{n,2n-k}$ also    splits  into  two real eigenvalues and \textit{vice versa}.  Therefore, in order to examine stability  of small-amplitude nonlinear modes that belong to the $n$th family,
it is sufficient to analyze only $n$
positive double eigenvalues $\Omega_{n,k}$  with
 $k=0, \ldots, n-1$.

In order to examine splitting of the double  eigenvalues $\Omega_{n,k}$, we employ Eqs.~(\ref{eq:expans}) which yield the  following asymptotic expansion for the linear stability operator:
${L} =  \tilde{{ L}}_n + \sigma \vep^2 {L}_n^{(2)} +
o(\vep^2)$, where  the correction ${L}_n^{(2)}$ is given as
\begin{equation*}
    \left(\!\!%
     \begin{array}{cc}%
    -b_n^{(2)} +  (\cK |\tw_n|^2) + \tw_n \cK_{\tw^*_n} &  \tw_n \cK_{\tw_n}\\%
      -  \tw_n^*\cK_{\tw_n^{*}} & b_n^{(2)} -  (\cK |\tw_n|^2) -   \tw_n^*\cK_{\tw_n}%
    \end{array}\!\!%
    \right).
\end{equation*}
Following the standard arguments of the perturbation theory for
linear operators~\cite{Kato}, in order to address  the splitting   of
the   double eigenvalue $\Omega_{n,k}$ we consider a $2\times2$
matrix defined as
\begin{eqnarray*}
     {M}_{n,k} = \left(
     \!\!
     \begin{array}{cc}%
    \displaystyle\frac{\langle  { L}_n^{(2)}\bp^{(I)}_{n,k}, {\bp^{(I)*}_{n,k}}\rangle}{\langle\bp^{(I)}_{n,k}, {\bp^{(I)*}_{n,k}}\rangle}&%
    \displaystyle\frac{\langle  { L}_n^{(2)}\bp^{(II)}_{n,2n-k}, {\bp^{(I)*}_{n,k}}\rangle}{\langle\bp^{(I)}_{n,k}, {\bp^{(I)*}_{n,k}}\rangle}\\[6mm]%
    \displaystyle\frac{\langle  { L}_n^{(2)}\bp^{(I)}_{n,k}, {\bp^{(II)*}_{n,2n-k}}\rangle}{\langle\bp^{(II)}_{n,2n-k}, {\bp^{(II)*}_{n,2n-k}}\rangle}&%
    \displaystyle\frac{\langle  { L}_n^{(2)}\bp^{(II)}_{n,2n-k}, {\bp^{(II)*}_{n,2n-k}}\rangle}{\langle\bp^{(II)}_{n,2n-k}, {\bp^{(II)*}_{n,2n-k}}\rangle}
    \end{array}
    \!\!
    \right),
\end{eqnarray*}
where $\langle {\bf a}, {\bf b}\rangle=\int {\bf b}^\dag(x) {\bf
a} (x)dx$ for any two column vectors ${\bf a}$ and ${\bf b}$ (here $^\dag$ means the transposition and complex conjugation). If
both the eigenvalues of   matrix  ${M}_{n,k}$ are real, then the double eigenvalue $\Omega_{n,k}$ splits into two real and simple eigenvalues. If such a situation takes
place for all $k=0,1,\ldots, n-1$, then  the small-amplitude
nonlinear modes $w_n(x)$ from the $n$th family are stable.  On the other hand, if  for some $k$ the
matrix ${M}_{n,k}$ has a complex eigenvalue, then the double
eigenvalue $\Omega_{n,k}$ splits into a pair of  complex-conjugated eigenvalues. This  is sufficient to conclude that  the small-amplitude
nonlinear modes from   the $n$th family are unstable.

Recall that real and imaginary parts of $\tw_n(x)$ have opposite parities. Using this fact, one can establish that  all the entries of   the matrices ${M}_{n,k}$   are real numbers which   read
%
%
\begin{eqnarray}
\label{eq:Mnk}
\left( {M}_{n,k}\right)_{1,1} = -b_n^{(2)} + \hspace{4cm}\nonumber\\%
\frac{\int \tw_k\{\tw_k \cK_{\tw_n^*} \tw_n   + \tw_n \cK_{\tw_n^*}\tw_k \}\ dx}{\int \tw_k^2dx},\\[2mm]
\left( {M}_{n,k}\right)_{2,2} = b_n^{(2)} - \hspace{4.3cm}\nonumber\\%
 \frac{\int\tw_{2n-k}\{\tw_{2n-k}\cK_{\tw_n^*}\tw_n+\tw_n\cK_{\tw_n^*}\tw_{2n-k} \}\ dx}{\int \tw_{2n-k}^2dx},\\[3mm]
\left({M}_{n,k}\right)_{2,1} = -\frac{\int\tw_n\tw_{2n-k}\cK_{\tw_n}\tw_k^*\  dx}{\int
\tw_{2n-k}^2dx},\\[3mm]
\left(  {M}_{n,k}\right)_{1,2} = \frac{\int\tw_n\tw_k \cK_{\tw_n}\tw_{2n-k}^*\ dx}{\int
\tw_{k}^2dx}.
\end{eqnarray}
Thus  in order to check  reality of the eigenvalues of the matrix ${M}_{n,k}$,  it is sufficient to consider the discriminant of its quadratic characteristic equation, i.e., the quantity $D_{n,k} = \left[({M}_{n,k}\right)_{1,1} +  \left({M}_{n,k}\right)_{2,2}]^2- 4\det {M}_{n,k}$. For $D_{n,k} > 0$ the eigenvalues of ${M}_{n,k}$  are   real  while  a situation with $D_{n,k}<0$ corresponds to a pair of complex conjugated eigenvalues. The situation $D_{n,k}=0$ requires a more delicate analysis as in this case behavior of the double  eigenvalue  $\Omega_{n,k}$ is determined by the next terms of the asymptotic expansions.

Using analytical expressions for the eigenfuctions $\tw_n(x)$,   the entries of  matrices ${M}_{n,k}$ and hence the determinant $D_{n,k}$ can be computed analytically  with a computer algebra  program (at least, for several relevant shapes of the kernel function $K(x)$). The resulting analytical expressions are too bulky to be presented in the   paper. However, for an interested reader to have an idea of how the expressions for the determinants $D_{n,k}$ read, in Appendix~\ref{app:Dnk} we provide two expressions for  $D_{1,0}$  and  $D_{2,0}$  obtained  for the Gaussian kernel and $\alpha=0$.  Complexity of the   expressions for $D_{n,k}$ increases drastically if larger numbers $n$ or nonzero values of $\alpha$ are involved.  Nevertheless, these  expressions  are available analytically and  can be used for finding the  domains of positivity and negativity of   the determinants $D_{n,k}$ which are  considered as functions of     $\lambda$ and $\alpha$.


\section{Results for the real parabolic potential}
\label{sec:cons}
Let us start presentation of our results with an important particular case $\alpha=0$, when
$\PT$-symmetric contribution is absent, and the model is
conservative.  First, it is relevant to notice that for $\alpha=0$
the eigenvalue problem (\ref{BdG}) admits an exact solution with
eigenvalue $\omega=2$ and the respective eigenvector given as
\begin{equation}
\label{eq:ex}%
\bp = \left(%
\begin{array}{c}%
\phantom{+}xw_n + w_n'\\%
-xw_n + w_n'%
\end{array}%
\right).
\end{equation}
Notice however that this exact solution does not hold
for nonzero $\alpha$.

Results of linear stability analysis of small-amplitude nonlinear modes from several first families  for the Gaussian kernel (\ref{eq:gauss}) are summarized in Table~\ref{tbl:al=0}. For
$n=1,2, \ldots 6$ and for $k=0, 1, \ldots, n-1$, Table~\ref{tbl:al=0}  reports sign of
the determinant $D_{n,k}$ in the local limit  $\lambda=0$ 
and the
approximated values of positive zeros of  determinants $D_{n,k}$  being considered as functions of $\lambda$ [at the same time, those are  the values of $\lambda$ at which functions $D_{n,k}=D_{n,k}(\lambda)$  change
their signs].    From Table~\ref{tbl:al=0} one can    recover information on  stability or instability of the
small-amplitude nonlinear modes   bifurcating from the $n$th linear
eigenstate for the given value of the nonlocality parameter $\lambda$: the  stability   takes place  if   $D_{n,k}(\lambda)>0$  for each $k=0,1,
\ldots, n-1$.

 It follows
from   Table~\ref{tbl:al=0} that for $n=1$ and  $k=0$ the
respective determinant $D_{1,0}(\lambda)$ is positive for any  $\lambda \geq 0$ (see also the explicit expression (\ref{eq:D10}) for  $D_{1,0}$ in Appendix~\ref{app:Dnk}).
Therefore, the small-amplitude nonlinear modes bifurcating   from the linear eigenstate $n=1$
are stable    for any  $\lambda \geq 0$.  The small-amplitude modes with  $n=2$ are unstable for $0\leq \lambda \lesssim 1.82$ but become stable for $\lambda \gtrsim 1.82$.  For $n>2$ the stability
situation might be   more complex with  the
half-axis $\lambda\geq 0$   divided into several alternating  intervals of stability
and instability.
In the local case   ($\lambda=0$)   all the  higher-order small-amplitude mode are unstable for $n\geq 2$ \cite{AlZez}. However, from Table~\ref{tbl:al=0} one observes that for each $n$ there exists a threshold value $\Lambda_n$
such that \textit{the small-amplitude modes become stable for the sufficiently  strong  nonlocality} $\lambda > \Lambda_n$. The
specific values of $\Lambda_n$ can be recovered  from
Table~\ref{tbl:al=0}: $\Lambda_1=0$, $\Lambda_2\approx 1.82$,
$\Lambda_3\approx 2.36$, etc. Thus  the sufficiently strong nonlocality essentially alters (enhances) stability of the excited modes.

We additionally mention that for each $n$ one has $D_{n, n-1}(\lambda)>0$ for all $\lambda$.
This is a consequence of the exact solution (\ref{eq:ex}) which implies that
the spectrum of operator $L$ always contains  an eigenvalue equal
to $2$. Therefore, the double  eigenvalue $\Omega_{n, n-1}=2$ always splits in the pair of real eigenvalues (one of which is equal to $2$). As a result,  the    eigenvalue $\Omega_{n, n-1}$  can  not  cause instability of   small-amplitude
nonlinear modes.

\begin{table}
\caption{Results of linear stability analysis for $\alpha=0$  and  the Gaussian kernel.}
\begin{tabular}{c@{\hspace{5mm}}c@{\hspace{5mm}}c@{\hspace{5mm}}c}
n & k & $\textrm{sign} \left.D_{n, k}(\lambda)\right|_{\lambda=0}$ &  zeros of $D_{n,k}(\lambda)$ (if any)\\[2mm]
\hline
1 & 0 & $+$ & no zeros\\[2mm]
2 & 0 & $-$ &  1.82\\
  & 1 & $+$ & no zeros\\[2mm]
3 & 0 & $-$ & 2.36\\
  & 1 & $-$ & 2.04\\
  & 2 & $+$ & no zeros\\[2mm]
4 & 0 & $+$ & 2.61 \quad 2.85\\
  & 1 & $-$ & 2.64 \\
  & 2 & $-$ & 2.26 \\
  & 3 & $+$ & no zeros \\[2mm]
5 & 0 & $+$ & 3.23 \quad 3.25\\
  & 1 & $-$ & 3.12 \\
  & 2 & $-$ & 2.90 \\
  & 3 & $-$ & 2.45 \\
  & 4 & $+$ & no zeros\\[2mm]
6 & 0 & $+$ & 3.566 \quad 3.568 \\
  & 1 & $+$ & 3.45 \quad 3.51 \\
  & 2 & $-$ & 3.37 \\
  & 3 & $-$ & 3.13 \\
  & 4 & $-$ & 2.64 \\
  & 5 & $+$ & no zeros
\end{tabular}
\label{tbl:al=0}
\end{table}

Let us now turn to examination of other   kernel functions $K(x)$. The results of linear stability analysis for the exponential kernel (\ref{eq:exponential}) are reported in Table~\ref{tbl:exp} which features an essentially different stability picture comparing to that for the Gaussian kernel. Specifically, we observe that for the exponential kernel  in the limit of strong nonlocality $\lambda\to\infty$, only the small-amplitude modes with $n=0, 1, 2$ are stable, whereas the modes with $n\geq 3$ are unstable both in the  local limit $\lambda=0$ and  for all positive  $\lambda$.

To summarize the  intermediate results obtained so far, we have encountered different stability situations for two different shapes of the kernel $K(x)$. In the case of the smooth Gaussian kernel, small-amplitude nonlinear modes bifurcating from the $n$th linear state become stable in the limit of high nonlocality for any $n$. On contrary, for the exponential kernel the small-amplitude modes are stable in the strongly nonlocal limit only for $n \leq 2$ and unstable for $n\geq 3$. The found difference in  stability  situation of strongly nonlocal
modes reveals certain correspondence  to the previous results on the
stability of the high-order nonlocal  nonlinear states, such as
multipole-mode 1D solitons \cite{XuKart05} and 2D spatial solitons
\cite{Skupin06}. Both \cite{XuKart05}  and \cite{Skupin06} report
that stable higher-order states can be found in the case of the
Gaussian kernel, while the exponential kernel results in
instability of all high order solitons, starting from some
threshold order. It has been also suggested in \cite{Skupin06}
that the mentioned difference between Gaussian and exponential
kernels is related to the singularity which the exponential kernel
has at the origin $x=0$. Our approach allows to substantiate the
above suggestion. Indeed, for  the Gaussian kernel one can  establish    asymptotic
behavior of entries of the matrix $M_{n,k}$ in the limit $\lambda \to \infty$. Specifically, for $\lambda
\to \infty$ one has
\begin{eqnarray}
\label{eq:asymp1}
(M_{n,k})_{1,1}, (M_{n,k})_{2,2}, &\propto& \lambda^{-3},\\
\label{eq:asymp2}
(M_{n,k})_{2,1}, (M_{n,k})_{1,2} &\propto& \lambda^{-(2n-2k+1)}.
\end{eqnarray}
Technical details on  computing the asymptotics (\ref{eq:asymp1})--(\ref{eq:asymp2}) are presented in Appendix~\ref{app:as}. Additionally, in Appendices~\ref{app:F} and \ref{app:HG} we provide  auxiliary relations necessary for   the derivation of (\ref{eq:asymp1})--(\ref{eq:asymp2}).

Relations (\ref{eq:asymp1})--(\ref{eq:asymp2}) immediately imply that  in the limit of large $\lambda$ for $k=0, 1,\ldots, n-2$ the diagonal elements of the matrix $M_{n,k}$ dominate the off-diagonal elements, which guarantees reality of the eigenvalues of  $M_{n,k}$.  Since the instability never occurs for $k=n-1$  [due to the presence of the exact solution (\ref{eq:ex}), as discussed above], we  confirm that for the Gaussian kernel (or, more generally, for an infinitely differentiable kernel)  the  small-amplitude modes of arbitrarily high  order $n$  are   stable in the limit of the strong  nonlocality.

At the same time, asymptotic results (\ref{eq:asymp1})--(\ref{eq:asymp2}), generically speaking, do not hold for a non-differentiable kernel. For example, for the exponential kernel we have computed that for $n=3$ and $k=0$  all the entries of the matrix $M_{n,k}$  behave as $\propto \lambda^{-2}$ for $\lambda\to\infty$, i.e., the  diagonal domination in the strongly nonlocal limit is not  guaranteed.

\begin{table}[tb]
\caption{Results of linear stability analysis for $\alpha=0$ and the exponential kernel.}
\begin{tabular}{c@{\hspace{5mm}}c@{\hspace{5mm}}c@{\hspace{5mm}}c@{\hspace{5mm}}}
n & k & $\textrm{sign} \left.D_{n, k}(\lambda)\right|_{\lambda=0}$ &  zeros of $D_{n,k}(\lambda)$ \\[2mm]
\hline
1 & 0 & $+$ & no zeros\\[2mm]
2 & 0 & $-$ &  1.31\\
  & 1 & $+$ & no zeros\\[2mm]
3 & 0 & $-$ & no zeros\\
  & 1 & $-$ & 1.54\\
  & 2 & $+$ & no zeros\\[2mm]
4 & 0 & $+$ & 7.25\\
  & 1 & $-$ & no zeros \\
  & 2 & $-$ & 1.70\\
  & 3 & $+$ & no zeros \\[2mm]
5 & 0 & $+$ & no zeros\\
  & 1 & $-$ & no zeros\\
  & 2 & $-$ & no zeros \\
  & 3 & $-$ & 1.84 \\
  & 4 & $+$ & no zeros\\[2mm]
6 & 0 & $+$ & no zeros \\
  & 1 & $+$ & no zeros \\
  & 2 & $-$ & no zeros \\
  & 3 & $-$ & no zeros \\
  & 4 & $-$ & 1.97 \\
  & 5 & $+$ & no zeros
\end{tabular}
\label{tbl:exp}
\end{table}

\begin{table}
\caption{Results of linear stability analysis for $\alpha=0$ and
the Lorentzian kernel.}
\begin{tabular}{c@{\hspace{5mm}}c@{\hspace{5mm}}c@{\hspace{5mm}}c@{\hspace{5mm}}}
n & k & $\textrm{sign} \left.D_{n, k}(\lambda)\right|_{\lambda=0}$ &  zeros of $D_{n,k}(\lambda)$ \\[2mm]
\hline
1 & 0 & $+$ & no zeros\\[2mm]
2 & 0 & $-$ &  1.52\\
  & 1 & $+$ & no zeros\\[2mm]
3 & 0 & $-$ & 2.70\\
  & 1 & $-$ & 1.66\\
  & 2 & $+$ & no zeros\\[2mm]
4 & 0 & $+$ & 2.77 \quad 3.446 \\
  & 1 & $-$ & 2.96 \\
  & 2 & $-$ & 1.77\\
  & 3 & $+$ & no zeros \\[2mm]
5 & 0 & $+$ & 3.88 \quad 3.991\\
  & 1 & $-$ & 3.74 \\
  & 2 & $-$ & 3.21 \\
  & 3 & $-$ & 1.87 \\
  & 4 & $+$ & no zeros\\[2mm]
6 & 0 & $+$ & 4.39 \quad 4.42 \\
  & 1 & $+$ & 3.98 \quad 4.28 \\
  & 2 & $-$ & 4.00 \\
  & 3 & $-$ & 3.44 \\
  & 4 & $-$ & 1.97 \\
  & 5 & $+$ & no zeros
\end{tabular}
\label{tbl:alg}
\end{table}

Table~\ref{tbl:alg} which reports the stability analysis for the algebraically depending kernel (\ref{eq:alg}) further confirms the stability of   strongly nonlocal high-order small-amplitude modes    for an infinitely differentiable kernel.
Besides of this,    we have also performed the same linear stability analysis for several nonphysical but still illustrative  nonlocal kernels such as a rectangular kernel with finite support
\begin{equation}
\label{eq:rect}
K(x) = \left\{%
\begin{array}{cc}
\frac{1}{2\lambda} & \mbox{for } x\in(-\lambda, \lambda)\\%
0 &  \mbox{otherwise,}
\end{array}
\right.
\end{equation}
a triangular shaped kernel with finite support
\begin{equation}
\label{eq:tri}
K(x) = \left\{%
\begin{array}{cc}
\frac{x}{\lambda^2}  + \frac{1}{\lambda}& \mbox{for } x\in(-\lambda, 0)\\%
\frac{-x}{\lambda^2}  + \frac{1}{\lambda}& \mbox{for } x\in(0, \lambda)\\%
0 &  \mbox{otherwise,}
\end{array}
\right.
\end{equation}
and a smooth  exponentially decaying kernel
\begin{equation}
\label{eq:erf}
K(x) = \frac{e^{-1/4}}{4\lambda }\left(  \textrm{erfc}\left(\frac{-x}{\lambda} \right)e^{-x/\lambda} + %
 \textrm{erfc}\left(\frac{x}{\lambda} \right)e^{x/\lambda} \right)
\end{equation}
[the particular choice of the kernels (\ref{eq:rect})--(\ref{eq:erf}) is motivated  by convenience of analytical evaluation of the respective  convolution integrals].  Kernels   (\ref{eq:rect}) and  (\ref{eq:erf})  also feature  stabilization of all higher-order strongly nonlocal modes,  whereas for  the triangular kernel (\ref{eq:tri})
the  higher-order strongly nonlocal  modes are unstable. 
This  allows to conclude that the  main feature responsible for the stability   is not the very presence of the singularity of the kernel, but the location of the singularity: the rectangular kernel (\ref{eq:rect}) features two discontinuities at $x=\pm \lambda$ but its behavior is nevertheless   similar to that of  the Gaussian and Lorentzian. The singularity at $x=0$ (the exponential and triangular kernels) does not allow for the stabilization of the higher-order small-amplitude modes.

\section{Results for the $\PT$-symmetric parabolic potential}
\label{sec:PT}
The linear stability analysis conducted above  can be naturally extended on the case of nonzero $\PT$-symmetric modulation $\alpha>0$. In this case, the  stability results can be conveniently represented in the plane $(\alpha, \lambda)$. Then the axis $\alpha=0$ recovers the results of the previous section, whereas   the horizontal axis $\lambda=0$ corresponds to the case of local $\PT$-symmetric parabolic potential  studied in \cite{ZezKon12}. The origin, i.e., the point $\lambda=\alpha=0$ is ``the most standard'' case of the real potential with the local nonlinearity \cite{AlZez}.

\begin{figure}
\includegraphics[width=\columnwidth]{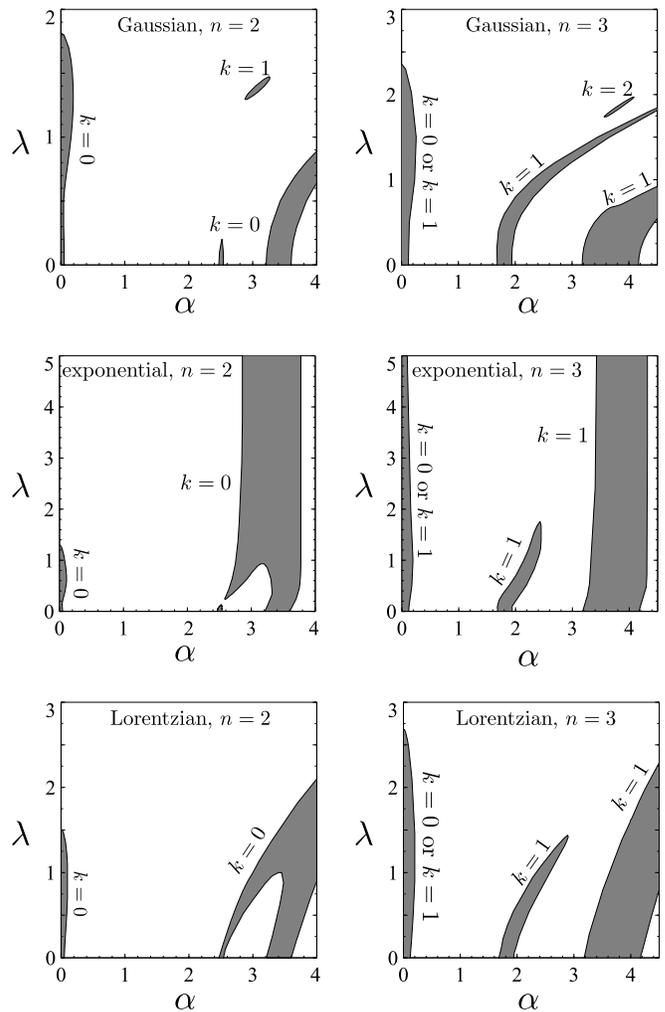}%
\caption{Linear limit stability diagrams  for $n=2$ and $n=3$ in the plane $(\alpha, \lambda)$. Darker domains correspond to instability.
Labels ``$k=0$'', ``$k=1$'', etc. at the unstable domains show  the number of the double eigenvalue $\Omega_{n,k}$ responsible for the instability.}
\label{fig:stdiagr}
\end{figure}

Similar to the previous section, we   focus  our attention  on the     Gaussian,  exponential and Lorentzian kernels. According to results of our analysis, for $n=1$ the small-amplitude nonlinear modes are stable for any values of $\lambda$ and $\alpha$ from the considered range. 
For the   families with $n\geq 2$, however, the stability situation becomes more complex and the  $(\alpha, \lambda)$-diagrams turn out  to be divided into the domains of stability and instability. Such stability diagrams for the families with $n=2$ and $n=3$ are presented in Fig.~\ref{fig:stdiagr}.

Turning to the local  limit ($\lambda=0$), one can observe that for
sufficiently large $\alpha$ the small-amplitude modes become
stable. Notice also
that  in the limit $\lambda \to 0$ all the considered kernels tend to the delta-function $\delta(x)$ and therefore display
the identical behavior. However, in accordance with results of the previous subsection, the considered kernels  feature  different behavior  in
the limit $\alpha=0$. As one can see in the  panels of the first and the third rows,  for the
Gaussian and Loretzian kernels, in the limit $\alpha=0$ large values of $\lambda$
lead to stabilization of the modes both for $n=2$ and $n=3$. For
the exponential kernel (panels in the second row)  the stabilization occurs for $n=2$ only,
while for $n=3$ the modes remain unstable for any arbitrarily large $\lambda$. However, this situation   changes drastically when a nonzero $\PT$
symmetry $\alpha>0$ is brought into consideration: say,  already for $\alpha=0.5$ the
nonlinear modes for the exponential kernel and $n=3$ become  stable for any $\lambda$ (from the considered range).  Overall, we can conclude that the diagrams presented in Fig.~\ref{fig:stdiagr} feature nontrivial and fairly interesting structure. In particular, one can observe that in some cases increase of $\alpha$ can lead either to stabilization or destabilization of the modes. The same remains true if one considers increase or decrease of $\lambda$. The stability diagrams allow to conjecture that   for any given $\lambda$ the small-amplitude modes eventually become stable in the limit $\alpha\to \infty$. However, in order to  verify  this conjecture, an additional study must be undertaken.


Notice also that for the case $\alpha>0$ allows for instability
caused by the eigenstate with $k=n-1$ (i.e., $k=1$ for $n=2$ and
$k=2$ for $n=3$). No such instabilities is  possible in the case
$\alpha=0$ due to the exact solution (\ref{eq:ex}).

\section{Nonlinear modes of finite amplitude}
\label{sec:fam}
The   stability results of  the previous sections  apply to nonlinear modes of infinitesimally small amplitude. One can check persistence of the   stability results by constructing numerically the families of nonlinear modes of finite amplitude and computing the eigenvalues of the corresponding linear stability problem. The families of nonlinear modes can be  visualized in the form of continuous curves on the plane $\sigma U$ \textit{vs.} $b$ where $U=\int |w_n(x)|^2dx$ is the total  energy flow. Then the small-amplitude modes are situated in the vicinity of the points $b_n = -(2n+1)$, $\sigma U = 0$.

\begin{figure}
\includegraphics[width=\columnwidth]{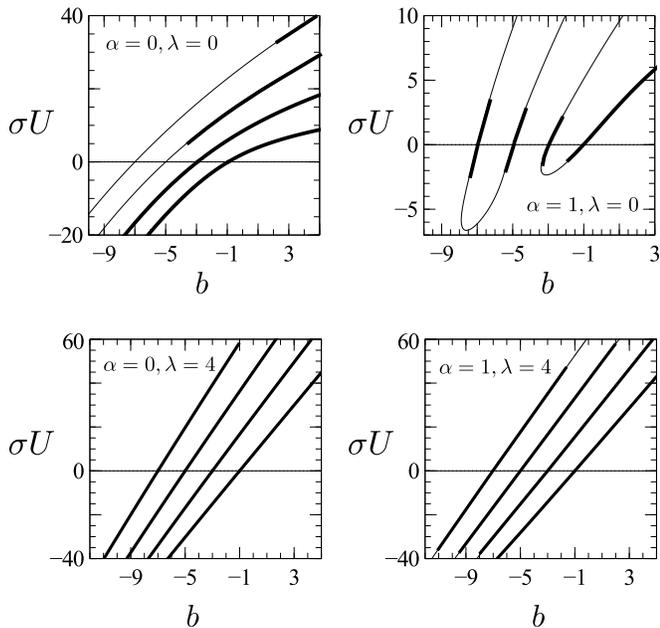}\\
\caption{The families of nonlinear modes with $n=0,1,2,3$ for different $\alpha$ and $\lambda$ (the Gaussian kernel). Stable modes correspond to the bold fragments of the curves.}
\label{fig:fams}
\end{figure}

Let us first recall the results related to the local case ($\lambda=0$). The corresponding diagrams for the conservative case \cite{AlZez} and for the $\PT$-symmetric case \cite{ZezKon12} are presented in the two upper panels of  Fig.~\ref{fig:fams}. The panels show  several families that bifurcate from the lowest linear eigenvalues $n=0,1,2,3$. In the conservative case, only the two lowest  families ($n=0,1$) are stable in the small-amplitude limit, whereas the small-amplitude modes with $n=2,3$ are unstable, In the $\PT$-symmetric case, all the four shown families are stable in the linear limit, but at least three of the four families loose stability as the strength of nonlinearity becomes sufficiently large.

The diagrams for the nonlocal case $\lambda=4$ (with the Gaussian kernel) are shown in the lower panels of  Fig.~\ref{fig:fams}. According to Table~\ref{tbl:al=0} and Fig.~\ref{fig:stdiagr}, at the chosen value of the nonlocality parameter, the nonlinear modes with $n=0,1,2,3$ are stable in the small-amplitude limit. The numerical results   in Fig.~\ref{fig:fams} indicate that the stability persists at least for small and moderate strengths of the nonlinearity ($|\sigma U| \lesssim 40$). Notice also that in the  panel with $(\alpha=1, \lambda=4)$  the nonlocality cancels the merging between the subsequent families which can be observed in the panel with  $(\alpha=1, \lambda=0)$. We however emphasize that an accurate  description  of the limit of strong nonlinearity  $|\sigma U| \to \infty$ should be elaborated  by means of a proper   asymptotic technique. While in the local case such  studies  have been already initiated   \cite{CPK10,GP14}, analysis of the nonlocal $\PT$-symmetric case in the limit of strong nonlinearity remains the open issue at the moment.

\section{Conclusion}
\label{sec:Concl}

In this study, we have demonstrated that the combined effect of the parabolic potential, nonlocality and $\PT$ symmetry  results in a nontrivial stability picture for high-order small-amplitude nonlinear modes. The specific results of our work can be outlined as follows:

\begin{itemize}
    \item A set of continuous families of nonlinear modes exists in the nonlocal $\PT$-symmetric nonlinear Schr\"odinger equation. The spectrum of the corresponding linear eigenvalue problem is available in the analytical form, and thus the stability problem for the small-amplitude nonlinear modes can be reduced to the searching real roots of certain analytical expressions.


    \item In the conservative case, small-amplitude nonlinear modes of arbitrary order become stable for sufficiently strong nonlocality, provided that the kernel is a sufficiently smooth function in the vicinity of the origin. If the kernel feature a singularity at $x=0$, no stabilization of high-order nonlinear modes is observed.

    \item Both the degree of nonlocality and the strength of the $\PT$ symmetry can be used to manage stability of the small-amplitude modes.

    \item The above stability conclusions remain valid for nonlinear  modes of finite  amplitude.

\end{itemize}




\appendix

\section{Examples of expressions for $D_{n,k}$}
\label{app:Dnk}

Let us present expressions $D_{1,0}$ and $D_{2,0}$ computed for the Gaussian kernel and $\alpha=0$:
\begin{widetext}
 \begin{eqnarray}
 \label{eq:D10}
D_{1,0} &=& \displaystyle \frac{ 1}{4}\,{\frac { \left( 2\,{\lambda}^{2}+1 \right) ^{2}}{\pi \left( \lambda^2+2\right)^5}},\\
D_{2,0} &=& {\frac {1}{64}}
\,\frac{13968\,{{  \lambda}}^{16}-28032\,{{\lambda}}^{14}-584\,{{\lambda}}^{12}-177136\,{{  \lambda}}^{10}-
54255\,{{\lambda}}^{8}-7564\,{{\lambda}}^{6}+4306\,{{\lambda}
}^{4}-948\,{{\lambda}}^{2}-207}{\pi \, \left( {{\lambda}}^{2}+2\right)^{13}}.
 \end{eqnarray}
 \end{widetext}
Obviously, $D_{1,0}$ is positive for any $\lambda$, while $D_{2,0}$ is negative for $\lambda=0$, but becomes positive for sufficiently large $\lambda$. Using the standard numerical tools for searching the roots of  polynomials, we find that $D_{2,0}$  has two real zeros:  $\lambda \approx  \pm 1.82$. The positive one   is presented in the respective row ($n=2$, $k=0$) of Table~\ref{tbl:al=0}.

\section{Definition and some properties of the Fourier transform}
\label{app:F}
We define the forward and backward Fourier transforms   as
\begin{eqnarray}
g(s)  = \cF\{f\} = \int f(x)e^{isx} dx,\\
f(x) = \cF^{-1}\{g\} = \frac{1}{2\pi} \int g(s)e^{-isx} ds.
\end{eqnarray}
Then   the Plancherel theorem says that
\begin{equation}
\label{eq:Planch}
\int f_1(x) f_2^*(x) dx =  \frac{1}{2\pi}\int g_1(s)g_2^*(s) ds,
\end{equation}
where $g_{1,2}(s)=\cF\{f_{1,2}\}$;  and from  the convolution theorem one has
\begin{equation}
\label{eq:conv}
\cF\{h_2\} = \cF\{K\} \cF\{h_1\},
\end{equation}
where   $h_2(x) = \int K(x-x') h_1(x')\ dx'$.

Another well-known relation  expresses the $p$th central moment of a function via the  Fourier transform ($p=0,1,\ldots$):
\begin{equation}
\label{eq:centr}
\int x^p f(x)\ dx = (-i)^p \left.\frac{d^p}{ds^p} g(s)\right|_{s=0}.
\end{equation}

\section{Some properties of the Hermite--Gauss eigenfunctions}
\label{app:HG}

Hereafter we assume the conservative case $\alpha=0$. The Hermite--Gauss eigenfunctions $\tw_n(x)$ are defined by Eqs.~(\ref{eq:twn}), see \cite{AS} for definition of the Hermite and Laguerre polynomials. For $\alpha=0$ the   modes $\tw_n(x)$ are  real-valued and satisfy  the   normalization
\begin{equation}
\label{eq:normsimple}
\int_{-\infty}^\infty \tw_n^2(x)\ dx = 1.
 \end{equation}
 The eigenfunctions are mutually orthogonal, i.e., $\int \tw_n\tw_k\ dx =0$ for any $n$ and $k\ne n$. Moreover, for any $n$ and  $k<n$:
\begin{eqnarray}
\label{eq:orth1}%
\int  x^p \tw_n(x)\, \tw_k(x)\ dx = 0, \quad p=0, 1, \ldots, n-k-1,\\
\label{eq:orth2}%
\int  x^{n-k} \tw_n(x)\tw_k(x)\ dx =   \sqrt{\frac{n!}{k!\,2^{n-k}}}.\quad\quad
\end{eqnarray}
In particular,
\begin{equation}
\label{eq:orthpart}
\int  x \tw_n(x)\tw_{n-1}(x)\ dx = \sqrt{\frac{n}{2}}.
\end{equation}
It is also useful to notice that
\begin{equation}
\label{eq:x^2w}
\int  x^2 \tw_n^2(x)\ dx = n + \frac{1}{2}.
\end{equation}

\section{Asymptotics in the strongly nonlocal limit}
\label{app:as}
We assume the conservative case $\alpha=0$ and an infinitely differentiable kernel written down as
\begin{equation}
K(x)=\frac{1}{\lambda}K_0\left(\frac{x}{\lambda}\right),
\end{equation}
where function $K_0(x)$   is even, real,
 and infinitely differentiable for any $x$.  Then
 one can adopt the Taylor series
\begin{equation}
\label{eq:Taylor}
K(x) = \sum_{p=0}^\infty  \frac{K_0^{(p)}(0) x^p}{\lambda^{p+1}
p!}.
\end{equation}

For the sake of brevity, in what follows let us  use the notation
\begin{equation}
F_{n,k}(s) = \cF\{\tw_n\tw_k\} = \int  \tw_n(x)\tw_k(x)e^{isx}dx.
\end{equation}

\textbf{Asymptotics for $b_n^{(2)}$.}
We start from Eq.~(\ref{b2}),  use Eqs.~(\ref{eq:Planch}),  (\ref{eq:conv}), (\ref{eq:normsimple}), and (\ref{eq:Taylor}),  as well as reality of the function   $F_{nn}(s)$, and evaluate  $b_n^{(2)}$ as follows:
\begin{eqnarray*}
b_n^{(2)} = (2\pi)^{-1}\int \cF\{K\}   F_{n,n}^2(s)  \ ds =\\
\int  K(x)\cF^{-1}\{F^2_{n,n}(s)\}\ dx =\\
\sum_{p=0}^\infty 
\frac{K_0^{(p)}(0)}{\lambda^{p+1}p!}\int  x^p \cF^{-1}\{F_{n,n}^2(s)\}\ dx
=\\
\sum_{p=0}^\infty  \frac{K_0^{(p)}(0)}{\lambda^{p+1}p!}(-i)^p \left.\frac{d^p}{ds^p}F_{nn}^2(s)\right|_{s=0} 
\end{eqnarray*}
(in order to obtain the latter equality we also used (\ref{eq:centr})).

From the definition of $F_{n,n}(s)$   and from Eqs.~(\ref{eq:normsimple})--(\ref{eq:x^2w}), one can  establish  the following   expansion:
\begin{equation}
\label{eq:Fnn}%
F_{n,n}(s) = 1 - \frac{1}{2}\left(n+\frac{1}{2}\right)s^2 + o(s^2)_{s\to 0}.
\end{equation}
%
Therefore
\begin{eqnarray}
\label{eq:b2as}
b_n^{(2)} =  \frac{K_0(0)}{\lambda} +\frac{K''_0(0)(2n+1)}{2\lambda^{3}} + O(\lambda^{-4})_{\lambda\to +\infty}.
\end{eqnarray}

\textbf{Asymptotics for $(M_{n,k})_{1,1}$.}
Since the asymptotics for $b_n^{(2)}$ is known from   Eq.~(\ref{eq:b2as}), it is convenient to consider the quantity $(M_{n,k})_{1,1} + b_n^{(2)}$ which can be evaluated as
\begin{eqnarray*}
\sum_{p=0}^\infty  \frac{K_0^{(p)}(0)}{\lambda^{p+1}p!}(-i)^p \frac{d^p}{ds^p}\left(F_{n,n}(s)F_{k,k}(s) + |F_{n,k}(s)|^2\right)_{s=0}.
\end{eqnarray*}
Using (\ref{eq:Fnn}) and that
\begin{equation}
F_{n,k}(s) = is\int  x\tw_n(x)\tw_k(x)\ dx + o(s)_{s\to 0},
\end{equation}
one can find  the resulting asymptotics
\begin{eqnarray*}
 (M_{n,k})_{1,1}=  -\frac{K''_0(0)\left(n-k + 2(\int  x\tw_n\tw_k\ dx)^2\right) }{2\lambda^{3}} \\+ O(\lambda^{-4})_{\lambda\to+\infty}.
\end{eqnarray*}
The terms decaying as $\lambda^{-1}$ cancel each other.

For $k=0, 1, \ldots, n-2$  one has $\int  x\tw_n\tw_k\ dx=0$ and therefore
\begin{eqnarray*}
 (M_{n,k})_{1,1}=  -\frac{K''_0(0) (n-k)  }{2\lambda^{3}} + O(\lambda^{-4})_{\lambda\to+\infty},
\end{eqnarray*}
while for $k=n-1$  one can use Eq.~(\ref{eq:orthpart}) and obtain
\begin{eqnarray*}
 (M_{n,n-1})_{1,1}=  -\frac{K''_0(0) (n+1)  }{2\lambda^{3}} + O(\lambda^{-4})_{\lambda\to+\infty}.
\end{eqnarray*}

\textbf{Asymptotics for $(M_{n,k})_{2,1}$.} Following to the same ideas, we can write down
\begin{equation}
\label{eq:M21as}
(M_{n,k})_{2,1} =-
\sum_{p=0}^\infty  \frac{K_0^{(p)}(0)}{\lambda^{p+1}p!}(-i)^p \frac{d^p}{ds^p}\left(F_{n,2n-k}(s)F_{n,k}^*(s)\right)_{s=0}.
\end{equation}
From definition of functions $F_{n,k}(s)$ and $F_{n,2n-k}(s)$ it follows that in vicinity of $s=0$ they behave as
\begin{eqnarray*}
F_{n,k}(s)  = \frac{i^{n-k}}{(n-k)!}\sqrt{\frac{n!}{k!\, 2^{n-k}}}\ s^{n-k} +o(s^{n-k})_{s\to 0},\\
F_{2n-k,k}(s) = \frac{i^{n-k}}{(n-k)!}\sqrt{\frac{(2n-k)!}{n!\, 2^{n-k}}}\ s^{n-k} +o(s^{n-k})_{s\to 0}.
\end{eqnarray*}

 Therefore several first terms of the series (\ref{eq:M21as}) are  zero, and the first nonzero term corresponds to $p=2(n-k)$ (provided that  $K_0^{(2n-2k)}(0) \ne 0$). Specifically,
\begin{eqnarray*}
 (M_{n,k})_{2,1}=
 \frac{(-1)^{n-k+1}K^{(2n-2k)}_0(0)}{2^{n-k}((n-k)!)^2\lambda^{2n-2k+1}}\sqrt{\frac{(2n-k)!}{k!}}\\ +  O\left(\frac{1}{\lambda^{2n-2k+2}}\right)_{\lambda\to+\infty}.
\end{eqnarray*}

Finally we notice that proceeding in  exactly the same way, one can prove analogous behavior of the other two entries $M_{n,k}$. Namely, $(M_{n,k})_{2,2}$ decays as $\lambda^{-3}$, while $(M_{n,k})_{1,2}$ decays as  $\lambda^{-(2n-2k+1)}$ for $\lambda\to\infty$.

\acknowledgments

The author is grateful to G. L. Alfimov  for very helpful discussions during the work on this project. The author also  acknowledges support of the FCT (Portugal)
grants PTDC/FIS-OPT/1918/2012 and UID/FIS/00618/2013.

\end{document}